\documentclass[a4paper,fleqn,usenatbib]{mnras}

\usepackage{newtxtext,newtxmath}
\usepackage{bbold, dsfont}

\usepackage[T1]{fontenc}
\usepackage{ae,aecompl}

\usepackage{graphicx}	
\usepackage{amsmath}	
\usepackage{amssymb}	
\usepackage{mathrsfs}
\usepackage[separate-uncertainty = true,multi-part-units=single]{siunitx}
\DeclareMathOperator{\sign}{sign}

\title[]{The radial density profile of Saturn's A ring}

\author[Gr\"atz et al.]{
Fabio M. Gr\"atz,$^{1}$\thanks{E-mail: fgraetz@uni-potsdam.de}
Michael Seiler,$^{1}$
Martin Sei\ss,$^{1}$
Holger Hoffmann$^{1}$
and Frank Spahn$^{1}$
\\
$^{1}$Institute of Physics and Astronomy, University of Potsdam, Karl-Liebknecht-Str. 24-25, 14476 Golm, Germany}

\date{Accepted XXX. Received YYY; in original form ZZZ}

\pubyear{2019}

\begin{document}
\label{firstpage}
\pagerange{\pageref{firstpage}--\pageref{lastpage}}
\maketitle

\begin{abstract}
  In this work we model the radial density profile of the outer A ring of
Saturn observed with Cassini cameras \citep{Tiscareno2018Icar}. An
axisymmetric diffusion model has been developed, accounting for the
outward viscous flow of the ring material, and the counteracting inward
drift caused by resonances of the planet's large outer moons. It has been 
generally accepted that the 7:6 resonance of Janus \emph{alone}
confines the outer A ring which, however, was disproved by
\citet{tajeddine2017}. We show that the step-like density profile of
the outer A ring is predominantly defined by the discrete first-order
resonances of Janus, the 5:3 second-order resonance of Mimas, and the
overlapping resonances of Prometheus and Pandora.
\end{abstract}

\begin{keywords}
  planets and satellites: rings -- hydrodynamics -- diffusion -- celestial mechanics
\end{keywords}



\section{Introduction}

The Voyager and Cassini space probes revealed that the rings of Saturn show a large amount of internal dynamical structures including density waves, bending waves, circumferential gaps with sharp edges and propeller objects. On large spatial scales, the ring's dynamics are predominantly governed by viscous spreading counteracted by resonant gravitational perturbations caused by embedded objects or larger satellites orbiting the planet outside of the ring system.

These outer satellites exert a systematic torque 
onto the ring material at the locations of their mutual inner Lindblad resonances. Resonances occur  when the mean motions  $\Omega$ and $\Omega_s$ of the ring particles and the 
satellites fulfill the 
resonance condition $\Omega/\Omega_{s}=j/(j-\ell)$ with $j,\ell\in\mathds N$ and
$\ell$ denotes the order of the resonance \citep{Goldreich1980ApJ, Greenberg1983Icar, MeyerVernet1987Icar}. 
They cause a drift of the ring particles which is always directed away from the moon, thus, in case of inner Lindblad
resonances, it is directed towards Saturn.
Viscous diffusion of the ring counteracts the gravitational perturbation and causes the ring to spread radially outwards.
In the close vicinity ($j\gg1$) of the perturber, i.e. an embedded moonlet, the distance between
successive resonances becomes small (meaning, that the number of resonances per radial interval
increases) and they, therefore, start to overlap. In this radial range 
the angular momentum transfer can be described by a continuous torque density \citep{Goldreich1980ApJ}.

Numerical and theoretical studies over the last decades have shown, that bodies
embedded in protoplanetary disks or planetary rings create \emph{a)}
S-shaped density modulations, dubbed propellers, if their masses are below a certain threshold or \emph{b)} cause a gap around the entire circumference of 
the disk if the embedded bodies mass exceeds the threshold (see 
\citet{Spahn1989124}, \citet{sremvcevic2000density,sremvcevic2002density} for theoretical and \citet{seiss_05, srem_2007, LEWIS2009, seiss_19} for numerical studies). The gravitational disturber exerts a torque on nearby disk 
particles, especially where resonances overlap, pushing them away from their original radial location, thus depleting 
the disk's density in its vicinity and trying to sweep free a gap. Diffusive spreading of 
the disk material due to particle interactions counteracts the gravitational 
scattering and has the tendency to refill the gap.

Propeller objects were predicted by \citep{sremvcevic2000density,sremvcevic2002density} and later discovered by the cameras on board the Cassini spacecraft (\citet{Tiscareno:2006aa}, \citet{srem_2007}, \citet{2008AJ_1351083T}, and \citet{2010ApJ_718L_92T}). The ring moons Pan and Daphnis by far exceed the threshold mass needed to maintain circumferential gaps. They were discovered in the Encke and Keeler gap by \citet{Showalter:1991} and \citet{daphins_entdeckung}, respectively.

\citet{tajeddine2017} recently discussed the common misconception that the A ring is solely confined by the 7:6 resonance of Janus \citep{MeyerVernet1987Icar}. They showed that the \emph{combined effort} of multiple resonances caused by the satellites Pan, Atlas, Prometheus, Pandora, Janus, Epimetheus, and Mimas gradually decreases the outward directed angular momentum flux in the A ring, so that the 7:6 resonance of Janus is finally able to truncate the disk.

Recently, \citet{Tiscareno2018Icar} have analyzed Cassini ISS
(Imaging Science Subsystem) data with the result of the most detailed density profile of the A ring to date. They systematically
searched for radial structures by using the continuous wavelet transform
technique and calculated the surface density from the wavelet
signature of A ring spiral waves. \citeauthor{Tiscareno2018Icar}
explain that a radial density profile is sufficient to describe the
basic structure of the rings, arguing that nearly all the fine
structure in Saturn's rings is either azimuthally symmetric or in the
form of a tightly-wound spiral.

\citet{graetz2018} described the radial density profile of a gap caused by an embedded moon in a planetary ring with a diffusion 
equation that accounts for the gravitational scattering of the ring particles as 
they pass the moon and for the counteracting viscous diffusion that has the 
tendency to fill the created gap. \citeauthor{graetz2018} applied the model to the Encke and Keeler gap to
estimate the shear viscosity $\nu$ of the ring, and to conclude that tiny
icy satellites cannot be the cause for the numerous
gaps observed in the C ring and the Cassini division. \citet{graetz2019} extended the diffusion model for circumferential gaps in dense planetary rings to account for the angular momentum flux reversal \citep{BGT82} in order to model the extremely sharp Encke and Keeler gap edges.

In this article, we build on our axisymmetric diffusion model for gaps in dense
planetary rings \citep{graetz2018} and bring the work of
\citeauthor{tajeddine2017} and \citeauthor{Tiscareno2018Icar} together by
developing a diffusion model that accounts for the viscous outward spreading of
the A ring and the counteracting torques exerted by the numerous resonances with
the outer moons in order to model the density profile of the outer A ring
presented by \citet{Tiscareno2018Icar}.

The plan of this article is as follows: First, we briefly summarize the 
derivation of the diffusion equation that describes the viscous spreading of 
the ring material as derived by \citet{graetz2018} (Section \ref{sec:hydro}). 
Then, we explain why and to what degree the material is pushed inwards at inner Lindblad resonances 
(Section \ref{sec:res}). In Section \ref{sec:results}, we apply our model to the 
outer A ring to calculate a surface density profile which we compare to the
measurements by \citet{Tiscareno2018Icar}. Finally, we conclude our 
findings and discuss which of the numerous moons are predominantly responsible for the density 
profile of the outer A ring (see Section \ref{sec:discussion}).

\section{Model}

\subsection{Hydrodynamic description}
\label{sec:hydro}

\citet{1981ApJ...248L..83L, ward81}, \citet{sremvcevic2000density, sremvcevic2002density} developed hydrodynamic models to describe dense planetary rings. In this article we build upon these models and present the derivation of the diffusion equation for circumferential gaps in planetary rings following \citet{graetz2018}.
Let us briefly summarize the steps: 
We start with the balances for mass and momentum that are given by the 
continuity and the Navier-Stokes equation using the thin-disk approximation with vertically averaged quantities:

\begin{align}
  \partial_t\Sigma+\nabla\cdot\left(\Sigma\vec u\right)&=0 \label{equ:continuity}\\
  \Sigma\partial_t\vec u+\Sigma\left(\vec u\cdot\nabla\right)\vec u&=\Sigma\vec 
    G-\nabla\cdot\mathbf P+\Sigma \vec f_m \label{equ:navier-stokes}
\end{align}

where $\Sigma, \vec{u}=(u,v)$, and $\mathbf{P}$ denote the surface mass density, the
velocity, and the pressure tensor. Here, $\vec G$ and $\vec f_{m}$ are the inertial 
accelerations $\vec G=\left(2\Omega v+3\Omega^2x,-2\Omega u\right)$ and the acceleration
exerted by the moons onto the ring material. 
In order to derive the diffusion equation we reduce the azimuthal velocity to the 
leading Kepler term and assume the radial component to be much smaller, 
$\vec u=\left(u,-3/2\cdot\Omega x\right)$, where $x=a-a_s$ is the radial 
distance from the moon. Here $a$ and $a_s$ denote the semi-major axes of the ring particle and 
the moon. Assuming that 
$\partial_x v = -3\Omega/2\gg \partial_xu$, and neglecting scalar pressure and bulk 
viscosity the pressure tensor can be approximated by

\begin{equation}
  \mathbf P=\nu\Sigma\begin{pmatrix}
 0 & \frac 32\Omega\\
 \frac 32\Omega & 0
\end{pmatrix} \, .
\end{equation}

In the following, we are interested in radial structures ($\partial_y\to 0$): 
The azimuthal component of the Navier-Stokes equation can be rewritten in the 
form $\Sigma\partial_t v+\Sigma u\partial_xv=-2\Sigma\Omega u-\partial_x\mathbf 
P_{xy}+\Sigma f_{m,y}$ and thus with $v=-3/2\cdot\Omega x$ in the form

\begin{equation}
   \Sigma u=\frac 2\Omega\left(\Sigma f_{m,y}-\frac 32 \Omega\partial_x\nu
     \Sigma\right) \label{eq:radial_mass_flux} \, .
\end{equation}

The average acceleration in azimuthal direction exerted by a disturbing 
moon on a ring particle with semi-major-axis $a$ is 
$\langle f_{m,y}\rangle=\Omega/2\cdot \mathrm d a/\mathrm dt$ 
\citep[see][Equation (31)]{graetz2018}. 
Inserting this radial mass flux (Equation (\ref{eq:radial_mass_flux})) into the 
continuity equation (Equation (\ref{equ:continuity})), yields a diffusion equation 
with an additional flux term accounting for the gravitational scattering of the 
ring material by a moon and for the rings viscous diffusion, two counteracting 
physical processes that define the radial density profile.  

\begin{equation}
  \partial_t\Sigma+\partial_x\left(\Sigma \frac{\mathrm da}{\mathrm dt} -
    3\partial_x\nu\Sigma\right)=0\label{eq:non_lin_diff_eq_v2}
\end{equation}
Here, 
$\nu$ denotes the ring's shear viscosity.

\citet{Charnoz_2010aa} used a similar diffusion equation as part of a hybrid simulation in which the viscous spreading of Saturn's rings past the Roche limit gives rise to the planet's small moons, reproducing the moons' mass distribution and orbital architecture.

Next, we calculate the drift of the ring particles caused by isolated and overlapping 
resonances.

\subsection{Resonances}
\label{sec:res}
For commensurable orbital frequencies between an outer perturbing moon, 
$\Omega_s$, and a test particle, $\Omega$, the orbital frequencies share the 
integer ratio

\begin{equation}
  \frac{\Omega_s}{\Omega} = \frac{j-\ell}{j} \le 1 \, ,
\end{equation}

where $j$ and $\ell=k+1$ are integers and $k=0$ for first-order resonances and $k=1$
for second-order resonances \citep{Greenberg1977VA}. In the following we set
$j^\star = \ell - j$. For example, the Janus 4:3 resonance means $j = 4$ and
$j^\star = -3$ ($k = 0$) in the used notation. Note that $G,m_p,a_s$ and 
$a$ denote the gravitational constant, the mass of Saturn and the moon's and 
the test particle's semi-major axis. Further, the masses of Saturn, the moon and 
the test particle follow the relation $m_p\gg m_s\gg m$. 
In our study, we include the oblateness of Saturn. Therefore, beside the 
orbital frequency

\begin{align}
  \Omega^2(r) & = \frac{Gm_c}{r^3}
  \left(1+\frac{3}{2}J_2\left(\frac{R_c}{r}\right)^2-\frac{15}{8}J_4\left(\frac{R_c}{r}\right)^4
  + \right. \nonumber \\
  & \phantom{=} + \left. \frac{35}{16}J_6\left(\frac{R_c}{r}\right)^6 \right)
\end{align}

the orbital motion of the test particle is further described by the epicyclic
and vertical frequencies, which slightly differ from $\Omega$ resulting in a
precession of the test particle's Kepler ellipse in radial and vertical
direction. In the following, we will focus on Lindblad resonances, 
which involve the precession of the argument of pericenter

\begin{equation}
  \dot{\varpi} = \Omega - \kappa \, ,
\end{equation}

with the epicyclic frequency, $\kappa$, given by

\begin{align}
  \kappa^2(r) & = \frac{Gm_c}{r^3}
  \left(1-\frac{3}{2}J_2\left(\frac{R_c}{r}\right)^2+\frac{45}{8}J_4\left(\frac{R_c}{r}\right)^4
  - \right. \nonumber \\
  & \phantom{=} - \left. \frac{1755}{16}J_6\left(\frac{R_c}{r}\right)^6\right) \, .
\end{align}

The values of the gravitational moments of Saturn used in our calculations are taken from \citet{Jacobson_2006}. From the Newtonian equations of motion describing the 
gravitational accelerations between Saturn as the central planet, the perturbing 
moon and the test-particle we can formulate the disturbing function
\begin{equation}
R = Gm_s \sum F(a,a_s,e,e_s) \cos \left( j \lambda_s + j^\star \lambda - \varpi -
  k \varpi_s \right) = - \varPhi
\end{equation}
as the disturbing potential. This infinite series depends on the orbital 
elements of the moon and the test particle and results from an elliptic 
expansion \citep[see e.g.][]{Brouwer1961Book, Murray1999Book}.

Here, our studies will focus on first- and second-order Lindblad resonances. 
Assuming small eccentricities of the perturbing moon and the test particle, 
which in fact is a valid assumption for the outer A ring, the 
effect of the resonance on the test particle's orbital motion can be described 
by the Lagrangian perturbation equations \citep{Murray1999Book}:

\begin{align}
  \dot{\Omega}  &  = 3 \left|j^\star\right| \, B_d \, \Omega^2 \, e \, e_s^k 
    \sin \varphi \, , \\
  \dot{e}  &  = B_d \, \Omega \, e_s^k \sin \varphi - Qe \, , \\
  \dot{\varpi} & = \dot{\varpi}_\mathrm{obl} - \Omega B_d \, \frac{e^k_s}{e} \cos \varphi
  \, , \\
  \dot{\varphi}  &  =  j \Omega_s + j^\star \Omega - \dot{\varpi}_\mathrm{obl}
  - k \dot{\varpi}_s + \Omega B_d\, \frac{e^k_s}{e} \cos \varphi \, .
  \label{equ:lagrange_perturbation}
\end{align}

Here, $\varphi = j \lambda_s + j^\star \lambda - \varpi - k \varpi_s$ 
represents the phase difference to the commensurability and 

\begin{equation}
  \nu_d = j \Omega_s + j^\star \Omega - \dot{\varpi}_\mathrm{obl} - 
    k \dot{\varpi}_s \approx \frac{3}{2} \left|j^\star\right| \Omega_\mathrm{res} \left( 
    \frac{a-a_{res}}{a_{res}} \right)
\end{equation}

denotes the distance to the exact resonance \citep[see e.g.][]{Hedman2010AJ}. 
Further, we assume a linear damping in the eccentricity (see Equation
(\ref{equ:lagrange_perturbation})) which mimics the role of dissipation by the 
ring material \citep{Greenberg1983Icar}. The dimensionless resonance strength 
$B_d = - \frac{m_s}{m_p} \beta^\prime f(j,\beta^\prime)$, with $\beta^\prime=\frac{a}{a_s}$, depends 
on the distance to Saturn. The resonance strength $f(j,\beta^\prime)$ is a function 
of the Laplacian coefficients

\begin{equation}
  b_\gamma^{(j)}(\beta^\prime) = \frac2 \pi \int\limits_0^{\pi}\mathrm d\Theta\,
    \frac{\cos (\Theta  j)}{\left(\beta^{\prime 2}-2 \beta^\prime  \cos (\Theta
    )+1\right)^{\gamma }} \, .
\end{equation}

For first-order Lindblad resonance the resonance strength is given by 

\begin{equation}
  f_d(j,\beta^\prime) = -\frac{1}{2} \left(2 j b_{\frac{1}{2}}^{(j)} \left(
    \beta^\prime\right)+\beta^\prime\sign(a_\text{s}-r) \frac{d}{d\beta^\prime} 
    b_{\frac{1}{2}}^{(j)}\left(\beta^\prime\right)\right) \, ,
\end{equation}

while for second-order resonances it is given by 

\begin{align}
  f_d(j,\beta^\prime) & = \frac 14 \left[ \left(-2 + 6 j - 4 j^2\right) 
    b_{\frac12}^{(j- 1)}\left(\beta^\prime\right) + \right. \nonumber \\ 
    & \phantom{=} \left.  + \left(2 - 4 j\right) \beta^\prime \frac{d}{d\beta^\prime} 
      b_{ \frac 12}^{(j - 1)}\left(\beta^\prime\right) - \beta^{\prime 2} 
      \frac{d^2}{d{\beta^{\prime 2}}} b_{ \frac 12}^{(j - 1)}
      \left(\beta^\prime\right)\right] \, .
\end{align}

Next, we introduce the conjugate variables 
$\mathrm h = e \cos \varphi$ and $\mathrm k = e \sin \varphi$ \citep{Hedman2010AJ} and solve their time 
derivative for the equilibrium condition $\dot{\mathrm h}=\dot{\mathrm k}=0$ which 
results in 

\begin{align}
\mathrm h & = - \frac{B_d \, \Omega \, \nu_d}{\nu_d^2+Q^2} \\
\mathrm k & = \frac{B_d \, \Omega \, Q}{\nu_d^2 + Q^2} \, .
\end{align}

Substituting the equilibrium solution for $\mathrm k$ into $\dot{\Omega}$ and using 

\begin{equation}
  \dot{\Omega} = - \frac{3}{2} \frac{\Omega}{a} \dot{a}
\end{equation}

finally results in the radial drift caused by the resonance

\begin{equation}
  \dot{a} = - 2 \left|j^\star\right| a \left(\Omega \, B_d \, e^k_s\right)^2
  \frac{Q}{\nu_d^2 + Q^2} \, .
  \label{equ:discrete_drift}
\end{equation}

Thus, at the resonance location $a_{res}$ the ring material is pushed 
to smaller radial positions which reduces the outward migration of the 
ring material and thus also reduces the surface mass density.
For large values of the dissipation factor $Q$ the resonances are broad and potentially overlapping whereas for small $Q$, the resonances are thin and isolated spikes of height $1/Q$. Since the total
angular momentum transferred by one resonance
onto the ring is constant and is not depending on $Q$, we set $Q$ to a 
constant value of $10^{-8}$ s$^{-1}$ yielding sharp, isolated resonances as we are mainly
interested in the height of the step a resonance causes in the density profile. The dissipation factor $Q$ was introduced by \citet{Greenberg1983Icar} who showed that an equatorial shepherding satellite cannot exert a torque on a circular or symmetrically distorted ring unless the induced eccentric motion of the ring particles is damped by some process, providing the asymmetry that permits a torque.

For small distances to the perturbing moon, or for large $j\gg1$ respectively,
the resonances start to overlap. In this close vicinity to the moon, the 
contribution of each resonance is hard to distinguish and thus can be 
described by the average drift rate \citep[see][]{Goldreich1980ApJ} given by

\begin{equation}                                 
 \frac{\mathrm da}{\mathrm dt}= \frac \alpha{x^4}\sign(x)\quad\text{with}\quad 
   \alpha=\frac {\mathscr A_1^2}{18\pi}\Omega \left(\frac{m_\text{s}}{m_\text{p}}
   \right)^2 a_s^5  \label{eq:dadt_derived}.
\end{equation}
with $\mathscr A_1=6.7187$ \citep[see][]{gt82,Seiss2010}. Equations (\ref{equ:discrete_drift}) and (\ref{eq:dadt_derived}) are used to describe the drift of the ring material caused by the perturbing moons in Equation (\ref{eq:non_lin_diff_eq_v2}).

\subsection{Numerical solution}

The density profile of the A ring is calculated by numerically solving 
the stationary diffusion equation (\ref{eq:non_lin_diff_eq_v2}), where 
$\mathrm da/\mathrm dt$ is the sum over the drifts caused by all perturbing
moons: Discrete resonances are described by Equation
(\ref{equ:discrete_drift}), the overlapping resonances of Prometheus
and Pandora by Equation (\ref{eq:dadt_derived}).
Resonances of Pan and Daphnis are not included in the calculations for simplicity as
(in a first approximation) they pass torque that is transferred to
them from inner ring regions on to outer regions of the ring.

The masses and orbital elements of the considered moons have been
obtained from the SPICE kernels sat375.bsp and sat378.bsp using the
algorithm presented by \citet{Renner2006}.  The normalized surface
mass density is set to $\Sigma/\Sigma_0=1$ at the left boundary and we model the viscosity dependence on the
surface mass density with a power-law
$\nu=\nu_0\left(\Sigma/\Sigma_0\right)^\beta$.

\section{Results}
\label{sec:results}

\begin{figure*}
  \centering
  \includegraphics[width=\textwidth]{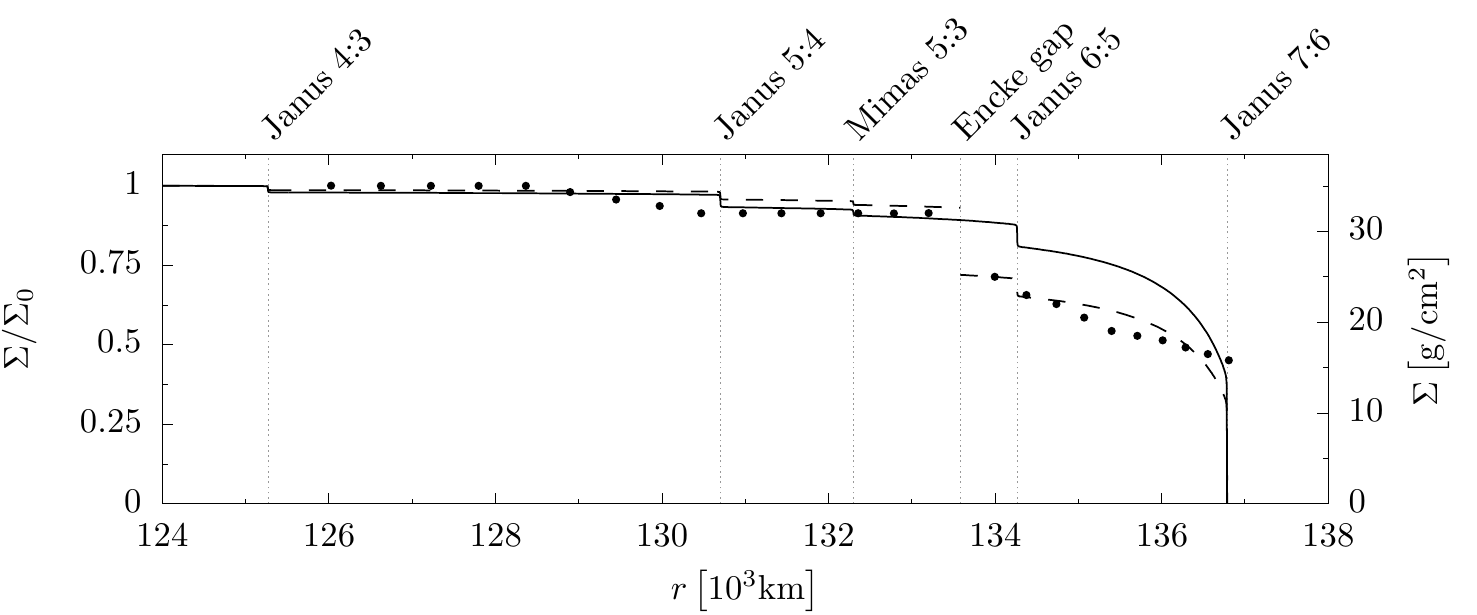}
  \caption{Density profile of the A ring. Dots are extracted from
  \citet{Tiscareno2018Icar}. The black solid line shows the solution of Equation 
  (\ref{eq:non_lin_diff_eq_v2}) for a viscosity of  
  $\nu_0=\SI{32}{\centi\meter\squared\per\second}$. However, this solution cannot 
  account for the abrupt drop of the surface mass density at the Encke gap. We 
  therefore included a second solution with 
  $\nu_0=\SI{49}{\centi\meter\squared\per\second}$ and a second boundary 
  condition of $\Sigma\approx 0.75\,\Sigma_0$ at the outside of the Encke gap 
  (black dashed line). The viscosities used to calculate the density profiles shown in this plot represent best-fit values.}
  \label{fig:profile}
\end{figure*}

Figure \ref{fig:profile} shows simulated and measured density profiles
of Saturn's A ring. The solid and dashed lines depict two solutions of
Equation (\ref{eq:non_lin_diff_eq_v2}) whose details are explained in
the following paragraphs. The black dots represent the surface mass
density measured by \citet{Tiscareno2018Icar} who derived the most
definitive surface density profile yet obtained for the A ring by
fitting the wavelet signature of spiral waves, estimating the error to
be at $\pm\SI{1}{\gram\per\centi\meter\squared}$.

The simulated density profile of the A ring illustrated by the solid
line in Figure \ref{fig:profile} assumes a viscosity of
$\nu_0=\SI{32}{\centi\meter\squared\per\second}$. The density profile
is dominated by first-order Lindblad resonances 4:3 to 7:6 of Janus,
the 5:3 second-order Lindblad resonance of Mimas, and the overlapping
resonances of Prometheus and Pandora: Inward of the Encke gap, the
model shows a stairway-like density profile caused by discrete
resonances which agrees reasonably well with the measurements. Outwards
of the 6:5 resonance of Janus, the surface density is reduced
continuously, not in steps, caused by the overlapping resonances of the
perturbing moons Prometheus and Pandora.  The aforementioned discrete
and overlapping resonances reduce the viscous outward flow of the ring
material so that the ring can be truncated at the 7:6 resonance of
Janus, causing the sharp outer A ring edge. Additional moons and
resonances -- such as Atlas (despite its vicinity to the A ring edge),
Epimetheus or second order resonances of
Janus -- do not significantly change the density profile.\\

At the location of the Encke gap, the measurements show a significant
drop in the surface mass density (inner edge
$\Sigma = \SI{32}{\gram\per\centi\meter\squared}$, outer edge
$\Sigma=\SI{25}{\gram\per\centi\meter\squared}$), which the model in
its current form (balance of the viscous outward flow and the inward
drift) cannot explain. However, if we presume the surface mass density
jump as the measurements show and define a second boundary condition
of $\Sigma\approx 0.75\,\Sigma_0$ at the radial location of the Encke
gap, then the model is able to reproduce the measured densities in the
trans-Encke-region fairly well. Different size distributions might be
responsible for the different surface mass densities and thus the
observed drop: \citet{FRENCH2000502} derived power-law particle size distributions for each of Saturn's main ring regions and found a lower size cutoff of $a_\text{min} \sim \SI{1}{\centi\meter}$ exterior to Encke gap and $a_\text{min} =\SI{30}{\centi\meter}$ interior to it. The trans-Encke region has a steeper power-law index than the region interior to Encke gap ($2.9$ and $2.75$, respectively). In addition, \citet{2014EGUGA..16.2479C} explained that the fraction of sub-centimeter particles increases in the outer A ring as the edge is approached. In theory, the jump in surface mass density at the Encke gap might even be of primordial origin.

The measurements show a nearly linearly decreasing surface mass
density outwards of the Encke gap whereas our simulations predict the
surface mass density to decrease like a concave function of the radial
distance to Saturn. This concave downwards decreasing surface mass
density is always found when one assumes a constant viscosity or
describes the viscosity with a power-law depending on $\Sigma$
\citep[see Figure 2 in][]{graetz2018}. This difference indicates that
describing the viscosity with a power-law is oversimplifying the $\Sigma$-dependence 
problem and that the real underlying physics is likely much more
complicated (see discussion in section \ref{sec:discussion}). Especially the size distribution should be expected to have a considerable influence on the transports.

\begin{figure*}
  \centering
  \includegraphics[width=\textwidth]{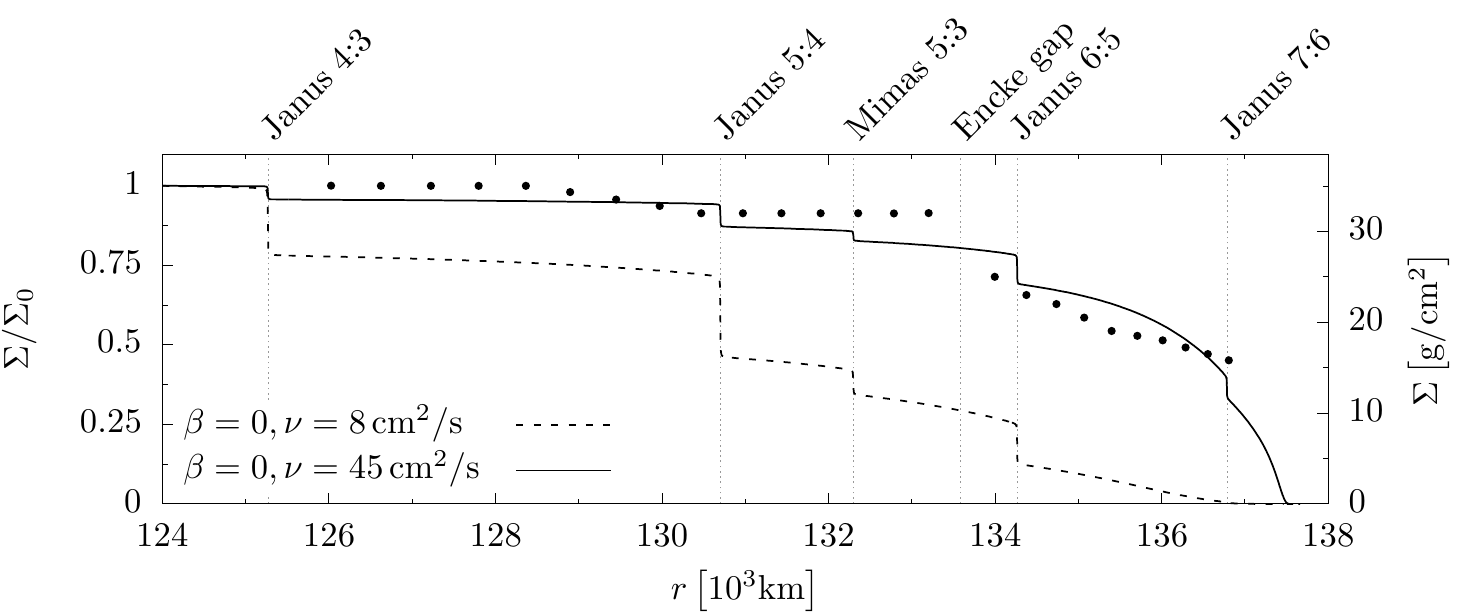}
  \includegraphics[width=\textwidth]{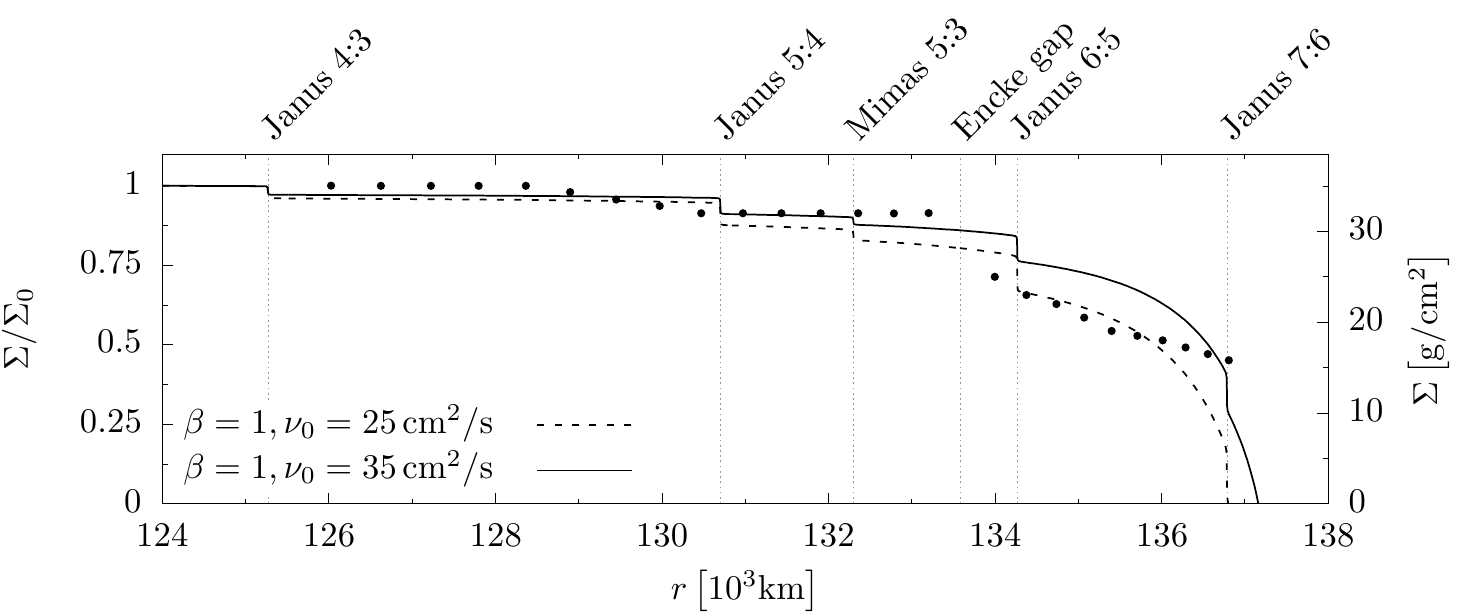}
  \caption{Solutions of the diffusion equation (\ref{eq:non_lin_diff_eq_v2}) for $\beta=0$ and $1$ (top and bottom panel, respectively). For the dashed curves the shear viscosity has been chosen so low that the 7:6 resonance with Janus can truncate the disk. The modeled surface mass densities are, however, significantly lower in the outer A ring than the measurements show. For any higher viscosities the material flows past the 7:6 resonance with Janus as exemplified by the solid curves, thus, contradicting the observations.}
  \label{fig3:beta01}
\end{figure*}

The power-law exponent of $\beta=2$ in the transport relation $\nu\sim\Sigma^\beta$ has been
predicted for the outer A ring by \citep{Daisaka2001} using N particle simulations. For
this and higher exponents our model works reasonably well while it
does not for lower values such as $\beta=0$ or $1$ as the
sharp edge of the outer A ring, as measured by
\citet{Tiscareno2018Icar} cannot be maintained. 
Figure \ref{fig3:beta01} illustrates why the sharp outer A ring edge, as measured by
\citet{Tiscareno2018Icar}, cannot be modeled using these power-law exponents:

\begin{enumerate}
\item The dashed lines have been obtained by choosing the maximum viscosities that still allow truncation of the ring at the 7:6 resonance with Janus. In the case of $\beta=0$ (upper panel), which corresponds to a constant shear viscosity, the surface mass density has decreased to approximately $0$ at the location of the 7:6 Janus resonance and is significantly lower than the measurements show throughout the entire central and outer A ring. In the case of $\beta=1$ (lower panel) the dashed curve models the surface mass density in the central A ring rather well but decreases to much lower values in the outer A ring than the measurements show ($\Sigma\approx 0.15\,\Sigma_0$ at the 7:6 resonance with Janus).
\item For any higher shear viscosities the viscous torque of the ring
  is larger than the counteracting torque of the 7:6 resonance with
  Janus. The disk is, thus, not
  truncated at the 7:6 resonance of Janus that defines the outer A
  ring edge and the material continues to flow outwards to be finally
  halted by Atlas (see black lines).
\end{enumerate}

The reason for this is as follows: The viscous torque
  $\Gamma_\nu$ is proportional to $\nu\Sigma$ \citep[Equation (54)]{MeyerVernet1987Icar} and thus, assuming a power-law for the viscosity, $\Gamma_\nu\propto \nu_0\Sigma^{\beta + 1}/\Sigma_0^\beta$. The 7:6 resonance with Janus that truncates the A ring exerts a certain torque; if in the model the viscous torque is larger, the material flows past the resonance, if it is smaller, the ring is truncated. As $\Gamma_\nu\propto \nu_0\Sigma^{\beta + 1}/\Sigma_0^\beta$, either $\Sigma$ can be decreased at the resonance location (by decreasing $\nu_0$) or $\beta$ can be increased in order to lower the viscous torque below the torque exerted by Janus. The surface mass density measurements by \citet{Tiscareno2018Icar} constrain how much the ring's surface mass density has decreased by the time the material reaches the outer A ring edge. For $\beta= 2$ the measurement can approximately be reproduced whereas for $\beta=0,1$ the surface mass density would have to be much lower than the measurements show for the 7:6 resonance to be able to truncate the disk.


\section{Summary and discussion}
\label{sec:discussion}

The main results of this article are:

\begin{enumerate}
\item The measured density profile by \citet{Tiscareno2018Icar} can
  be modeled surprisingly well with an axisymmetric diffusion model,
  if we assume the measured surface mass density drop of $\approx 25\%$ at
  the radial location of the Encke gap as a second boundary condition.
\item The 4:3 to 7:6 Lindblad resonances
  of Janus, the 5:3 Lindblad resonance of Mimas and the overlapping resonances
  of Prometheus and Pandora govern the density profile of the A ring and lower the surface mass
  density enough so that the 7:6 resonance of Janus can form and
  maintain the observed sharp outer A ring edge.
\item Resonances of Epimetheus are considerably weaker than Janus' resonances and do not alter the density profile significantly. Atlas -- despite its vicinity to the A ring edge -- has been found to have only very minor influence on the profile of the A ring because of its low mass as well.
\item The model favors high power law exponents, i.e. $\beta = 2$ as suggested by \citet{Daisaka2001}. The sharp outer A ring edge cannot be modeled for $\beta = 0,1$.
\end{enumerate}

The power-law we used to describe the shear viscosity's dependency on the surface mass density, which was introduced by \citet{schmit_tscharnuter_95}, is a reasonable simplification often used in hydrodynamic descriptions of planetary rings \citep{seiss2011}. Relating thereto we want to point out, that the goal of this article is to demonstrate that a fluid model using a simple description of the viscosity is able to describe the measured density profile surprisingly well but that the viscosities used to calculate the density profiles shown in Figure \ref{fig:profile} should be regarded as best-fit values for this model rather than measurements of the viscosity for the entire A ring. The model in its current form is not able to model the almost linear decrease in surface mass density in the trans-Encke region and in order to improve the model
one might have to take several additional aspects into consideration:

\begin{enumerate}
  \item Especially the particle size distribution should be expected to have a considerable influence on the transports.
\item In addition, the size distribution is expected to change radially, especially in strongly perturbed regions. \citet{2014EGUGA..16.2479C}, for instance, showed that the fraction of sub-centimeter particles increases in the outer A ring as the edge is approached. Parameters of the model such as $\nu_0$ and $\beta$ should, thus, be functions of $r$.
\item A better description of the trans-Encke region might require modeling more physics such as the ring's self gravity or aggregation and fragmentation.
\end{enumerate}

Even having accounted for the additional aspects just mentioned, a hydrodynamic description might prove to be too simple, so that a proper kinetic description might have to be used. We want to emphasize that the usage of transport coefficients such as the viscosity is always a linear approximation around equilibrium (Onsager) which cannot be expected to yield a perfect modeling of the observations, especially in as highly perturbed granular media as the outer A ring.

That being said, the values for the viscosities we used are in good agreement to the values \citet[Figure 1]{tajeddine2017} report: At the left boundary ($r=\SI{124e3}{\kilo\meter}$) we set the viscosity to  
$\nu_0=\SI{32}{\centi\meter\squared\per\second}$ and 
$\nu_0=\SI{49}{\centi\meter\squared\per\second}$ for the solid and dashed 
solution in Figure \ref{fig:profile} respectively. At this radial position 
\citet[Figure 1]{tajeddine2017} report an effective viscosity between  
$\nu_0=\SI{37}{\centi\meter\squared\per\second}$ and  
$\nu_0=\SI{48}{\centi\meter\squared\per\second}$.

While there are small inconsistencies to the measurements (see the
measured linear decreasing surface mass density compared to the modeled concave down
decrease for the trans-Encke region), this simple model describes the radial density profile of the A ring surprisingly well and we consider this simple approach as a 
step towards modeling the radial density profile of the A ring by
\citet{Tiscareno2018Icar} (semi-)analytically. In the future, it might prove to be very interesting to further analyze the viscosity's dependency on the surface mass density with a kinetic approach, especially in the limit of small, vanishing densities that occur at the outer A ring edge.

\section*{Acknowledgements}
This work has been financed by the Studienstiftung des deutschen Volkes, the 
Deutsche Forschungsgemeinschaft (Sp 384/28-2 and Ho5720/1-1), and the Deutsches Zentrum f\"ur 
Luft-und Raumfahrt (OH 1401).



\bibliographystyle{mnras}
\bibliography{refs} 




\bsp	
\label{lastpage}
\end{document}